\documentclass[11pt,a4paper]{article}
\usepackage{jheppub}
\usepackage[utf8]{inputenc}
\usepackage{latexsym, graphicx} 
\usepackage{amsmath,amsfonts,amssymb}
\usepackage{mathrsfs}
\usepackage{tensor}
\usepackage{caption,subcaption}

\newcommand*{\zbar}{\bar{z}}
\newcommand*{\wbar}{\bar{w}}

\newcommand*{\dz}{\partial_z}

\newcommand*{\partialbar}{\bar{\partial}}
\newcommand*{\Pibar}{\bar{\Pi}}

\newcommand*{\Oh}{\mathcal{O}_h}
\newcommand*{\Bh}{\mathcal{B}_h}

\title{Reparametrization modes in 2d CFT and\\
the effective theory of stress tensor exchanges}
\author[]{Kevin Nguyen}
\emailAdd{kevin.nguyen@kcl.ac.uk}
\affiliation[]{Department of Mathematics, King's College London, London, United Kingdom}

\abstract{We study the origin of the recently proposed effective theory of stress tensor exchanges based on reparametrization modes, that has been used to efficiently compute Virasoro identity blocks at large central charge. We first provide a derivation of the nonlinear Alekseev--Shatashvili action governing these reparametrization modes, and argue that it should be interpreted as the generating functional of stress tensor correlations on manifolds related to the plane by conformal transformations. In addition, we demonstrate that the rules previously prescribed with the reparametrization formalism for computing Virasoro identity blocks naturally emerge when evaluating Feynman diagrams associated with stress tensor exchanges between pairs of external primary operators. We make a few comments on the connection of these results to gravitational theories and holography.}

\begin{document}

\maketitle
\flushbottom

\section{Introduction}

An effective action used to describe stress tensor exchanges in conformal field theories (CFT) has been recently constructed by Haehl, Reeves and Rozali \cite{Haehl:2018izb,Haehl:2019eae}. In two dimensions, it reads 
\begin{equation}
\label{eq:Haehl}
W=-\frac{c}{192\pi} \int d^2x\, \epsilon^i \left(\delta_{ij} \square -2 \partial_i \partial_j\right) \square \epsilon^j + O(\epsilon^3),
\end{equation}
where $\epsilon^i(x)$ is known as the \textit{reparametrization mode}, and $\delta_{ij}$ is the flat euclidean background metric. The reparametrization mode is interpreted as the generator of the infinitesimal transformation
\begin{equation}
\label{eq:delta g intro}
\delta g_{ij}=\partial_i \epsilon_j + \partial_j \epsilon_i - \delta_{ij}\,  \partial_k \epsilon^k,
\end{equation}
resulting from a change of coordinate followed by a Weyl rescaling of the background metric. When $\epsilon^i(x)$ is a conformal Killing vector field, it is the symmetry parameter of a \textit{conformal transformation} and the effective action \eqref{eq:Haehl} correspondingly vanishes. By contrast, generic configurations $\epsilon^i(x)$ do not generate spacetime symmetries and acquire a nonzero action due to the Weyl anomaly. Hence one can think of $\epsilon^i(x)$ as a pseudo-Goldstone mode resulting from broken Weyl invariance. The effective action \eqref{eq:Haehl} admits a nonlinear extension known as the Alekseev--Shatashvili action \cite{Alekseev:1988ce}. Quite remarkably, Cotler and Jensen showed that this nonlinear extension describes pure gravity in three dimensions with anti-de Sitter (AdS) asymptotics~\cite{Cotler:2018zff}, whose symmetries are well-known to be that of a two-dimensional CFT \cite{Brown:1986nw}. Through this specific example, they therefore provided a completion of \eqref{eq:Haehl}. The first goal of the present work will be to explain the origin and role of Alekseev--Shatashvili action from a CFT perspective.

The primary interest of the reparametrization mode formalism is that it provides an efficient way to compute Virasoro identity blocks, i.e., to evaluate the contributions to correlation functions coming from stress tensor exchanges between pairs of external primary fields. To achieve this, one first introduces a set of \textit{bilocal vertex operators} that are closely related to reparametrized primary two-point functions. Within the reparametrization formalism, these bilocal operators are used as effective couplings between a pair of identical primary operators and the reparametrization mode itself. One can use them together with the dynamics dictated by the effective action \eqref{eq:Haehl} to compute Virasoro identity blocks. Agreement with previously known results at large central charge \cite{Fitzpatrick:2014vua,Fitzpatrick:2015zha,Fitzpatrick:2015dlt,Beccaria:2015shq} was found in \cite{Cotler:2018zff}, thereby demonstrating the utility of the reparametrization mode formalism. 
Finally, a partial justification of the above procedure was provided through the identification of the reparametrization mode with the shadow operator of the stress tensor \cite{Haehl:2019eae}.

The reparametrization mode formalism builds on a series of works aiming at a universal description of out-of-time-order correlators (OTOCs) in maximally chaotic quantum systems   \cite{Turiaci:2016cvo,Jensen:2016pah,Blake:2017ris,Haehl:2018izb,Cotler:2018zff,Jensen:2019cmr}. The exponential Lyapunov behavior displayed by OTOCs in such systems being universally controlled by their temperature \cite{Maldacena:2015waa}, it was therefore natural to look for a universal effective description in terms of reparametrization modes. Agreement was indeed found between this effective description and more conventional CFT methods~\cite{Haehl:2018izb}. Although this is not the primary focus of the present paper, the strong connection between maximal chaos and the AdS/CFT correspondence is worth mentioning as it appears that holographic CFTs are maximally chaotic \cite{Shenker:2013pqa,Shenker:2014cwa,Roberts:2014ifa,Mezei:2019dfv,Balasubramanian:2019stt,Craps:2020ahu,Poojary:2018esz,Jahnke:2019gxr}.

The aim of the present work is to clarify the origin of the reparametrization mode formalism. In particular, we provide a derivation of the nonlinear version of \eqref{eq:Haehl} from first principles, starting from the Polyakov action as the universal generating functional for stress tensor correlations in any 2d CFT \cite{Polyakov:1981rd}. We review basic properties of the Polyakov action in section~\ref{section:Polyakov}, and describe in section~\ref{section:effective action} how it reduces to the Alekseev--Shatashvili action when the background metric, considered as a source for the stress tensor, is generated from the flat metric by a finite version of \eqref{eq:delta g intro}. We show that this nonlinear extension of the effective action \eqref{eq:Haehl} should still be viewed as a generating functional for (holomorphic) stress tensor correlations on manifolds related to the plane by conformal transformations, with the derivative of the reparametrization mode acting as the corresponding source\footnote{A similar use of the reparametrization mode as a source for the stress tensor has been made in the context of $T\bar{T}$ deformations, however without reference to a nonlinear action for the reparametrization mode \cite{Hirano:2020ppu}.}. In section~\ref{section:exchanges}, we revisit the computation of Virasoro identity blocks in the reparametrization formalism. In particular, we show that the prescriptions given in \cite{Cotler:2018zff,Haehl:2019eae,Anous:2020vtw} naturally emerge when evaluating Feynman diagrams involving stress tensor exchanges between external pairs of identical primary operators. This provides a new justification for these otherwise mysterious prescriptions, independent from the earlier one based on the shadow operator formalism and originally presented in \cite{Haehl:2019eae}. We end with a discussion of the results, and point towards possible further developments and applications of the formalism described here. We also comment on connections to theories of gravity and holography.

\paragraph{Conventions.} We work in euclidean signature. We use the shorthand notations $T\equiv -2\pi T_{zz}$ for the holomorphic component of the stress tensor, $\delta(z)\equiv \delta^{(2)}(z,\zbar)$ for the delta distribution normalized as $\int d^2z\, \delta(z)=1$, and $z_{ij}\equiv z_i-z_j$ for relative distances. To avoid clutter, we sometimes suppress the functional dependence on coordinate labels, and write expressions such as $\partialbar \epsilon_1 \equiv \partial_{\zbar_1} \epsilon(z_1,\zbar_1)$. We make repetitive use of the magic distributional identity
\begin{equation}
\label{eq:magic}
\partial_{\zbar}\left( \frac{1}{z}\right)=2\pi \delta(z)\,,
\end{equation}
which will be our main computational weapon. 

\section{The Polyakov action}
\label{section:Polyakov}
The starting point for the construction of the effective action \eqref{eq:Haehl} in \cite{Haehl:2019eae} was the generating functional for the connected stress tensor two-point function on the complex plane,
\begin{equation}
W\left[\delta g\right]=\frac{1}{8} \int d^2x\, d^2y\, \delta g^{ij}(x)  \delta g^{mn}(y) \langle T_{ij}(x) T_{mn}(y) \rangle_{\text{plane}}+O(\delta g^3)\,,
\end{equation}
where the metric perturbation $\delta g^{ij}$ is a source for the stress tensor $T_{ij}$. 
Upon insertion of \eqref{eq:delta g intro}, it was shown to reduce to the effective action \eqref{eq:Haehl} in terms of the reparametrization mode $\epsilon^i$ \cite{Haehl:2019eae}. It is reasonable to expect that its nonlinear extension can be obtained by considering the generating functional of \textit{all} connected stress tensor correlators. This generating functional appears to be universal in two dimensions and has been derived in closed form long ago by Polyakov \cite{Polyakov:1981rd}, which we briefly review in this section.

Stress tensor correlation functions on the plane are fully constrained by conformal symmetry, the only dependence on a given theory occurring through the central charge $c$. Up to its value, the generating functional $W\left[g_{ij}\right]$ of all connected stress tensor correlators is therefore universal. Polyakov's starting point for its construction is the anomalous trace of the stress tensor expectation value on a space with arbitrary background metric $g_{ij}$ and curvature $R$,
\begin{equation}
\frac{c}{24 \pi} R=g^{ij} \langle T_{ij} \rangle=-\frac{2}{\sqrt{g}}\, g^{ij} \frac{\delta W}{\delta g^{ij}},
\end{equation}
where the last equality follows from the very definition of the generating functional $W\left[g_{ij}\right]$. Integrating this equation, Polyakov obtained \cite{Polyakov:1981rd}
\begin{subequations}
\label{eq:Polyakov action}
\begin{align}
W\left[g_{ij}\right]&=-\frac{c}{96\pi} \int d^2x\, d^2y\, \sqrt{g(x)}\, \sqrt{g(y)}\, R(x) G(x,y) R(y)\\
&=-\frac{c}{96\pi} \int d^2x\, \sqrt{g(x)}\, R(x) \frac{1}{\square} R(x),
\end{align} 
\end{subequations}
where $G(x,y)$ is the Green function solution to
\begin{equation}
\square G(x,y)=\frac{\delta^{(2)}(x-y)}{\sqrt{g(x)}}.
\end{equation}
The Polyakov action \eqref{eq:Polyakov action} is manifestly nonlocal in the background metric $g_{ij}$ that acts as a source for the stress tensor. It can be put in an alternative form through the introduction of an auxiliary variable $\phi$ solving
\begin{equation}
\label{eq:boxphi}
\square \phi=R,
\end{equation} 
such that the generating functional coincides with the action of a Liouville theory,
\begin{equation}
\label{eq:Liouville action}
W\left[g_{ij}\right]=-\frac{c}{48\pi} \int d^2x\, \sqrt{g(x)}\, \left(\frac{1}{2}(\partial \phi)^2+\phi R\right)\,.
\end{equation} 
We stress that the Liouville field $\phi$ is not an independent variable, but rather a nonlocal functional of the metric through \eqref{eq:boxphi}. The stress tensor expectation value may be computed from \eqref{eq:Liouville action} by functional differentiation, and is found to coincide with the classical Liouville stress tensor 
\begin{equation}
\label{eq:Tij}
\langle T_{ij} \rangle=-\frac{2}{\sqrt{g}} \frac{\delta W}{\delta g^{ij}}=\frac{c}{24\pi} \left[\frac{1}{2}\partial_i \phi\, \partial_j \phi-\nabla_i \nabla_j \phi+g_{ij} \left(\square \phi-\frac{1}{4} (\partial \phi)^2 \right)  \right]=T_{ij}^{\phi}\,.
\end{equation}
Consistently, one recovers the trace anomaly which we started from,
\begin{equation}
\label{eq:trace anomaly}
g^{ij} \langle T_{ij} \rangle =\frac{c}{24\pi}\, R\,.
\end{equation}
It is worth mentioning that covariance of \eqref{eq:boxphi} under a Weyl rescaling 
\begin{equation}
\label{eq:Weyl rescaling}
g_{ij} \mapsto e^{\omega} g_{ij}\,,
\end{equation}
implies that $\phi$ must transform by a shift $\phi \mapsto \phi- \omega$. It is therefore natural to interpret $\phi$ as the pseudo-Goldstone mode associated to broken Weyl symmetry. The expectation value \eqref{eq:trace anomaly}, or equivalently the configuration $\phi$ determined through \eqref{eq:boxphi}, labels one of the broken vacua. Due to explicit breaking of Weyl symmetry by the central charge $c$, this pseudo-Goldstone mode acquires a nonzero action \eqref{eq:Liouville action}. 

Functional differentiation of the generating functional $W[g_{ij}]$ yields connected stress tensor correlators on a background with fixed metric $g_0$,
\begin{equation}
\langle T_{ij}(x_1) ... T_{mn}(x_n) \rangle_{g_0} = \frac{(-2)^n}{\sqrt{g(x_1)} ... \sqrt{g(x_n)}}\, \frac{\delta^n W}{\delta g^{ij}(x_1) ... \delta g^{mn}(x_n)}\Big|_{g=g_0}+...\,,
\end{equation}
where the dots refer to contact terms resulting from functional differentiation of the metric determinant of the type
\begin{equation}
\frac{\delta}{\delta g^{ij}(x_k)}\left( \frac{1}{\sqrt{g(x_l)}}\right)\,.
\end{equation}
For instance, we can compute the stress tensor two-point function on the plane equipped with flat metric euclidean metric. A straightforward computation yields
\begin{equation}
\langle T_{ij}(x) T_{mn}(y) \rangle_{\text{plane}}=-\frac{c}{48\pi^2} \left(\delta_{ij}\square^x -\nabla_i^x \nabla_j^x \right) \left(\delta_{mn} \square^y -\nabla_m^y \nabla_n^y \right) \ln \mu^2|x-y|^2,
\end{equation}
where $\mu$ is an arbitrary energy scale introduced such that the argument of the logarithm is dimensionless. With complex coordinates
\begin{equation}
\label{eq:flat metric}
ds^2=dz\, d\zbar,
\end{equation} 
one recovers in particular the standard expression 
\begin{equation}
\langle T(z,\zbar) T(w,\bar{w}) \rangle= \frac{c}{2(z-w)^4}\,.
\end{equation}

\paragraph{Weyl non-invariance.}
The Polyakov action \eqref{eq:Polyakov action} arises from the breaking of Weyl invariance by quantum effects. It should therefore be expected that it transforms nontrivially under Weyl rescalings. Using
\begin{equation}
R\left[e^\omega g_{ij}\right]=e^{-\omega}\left(R_g-\square_g \omega\right), \qquad \det e^\omega g=e^{2\omega} \det g,
\end{equation}
one indeed finds that \eqref{eq:Polyakov action} transforms in a nontrivial way,
\begin{equation}
\label{eq:Gamma transformation}
W\left[e^\omega g_{ij}\right]=W\left[g_{ij}\right]+\frac{c}{48\pi} \int d^2x\, \sqrt{g(x)}\, \left(\frac{1}{2}(\partial \omega)^2+\omega R_g\right).
\end{equation}
As one could have anticipated from the previous discussion, the term spoiling Weyl invariance is precisely of the form of a Liouville action for the conformal factor $\omega$. 

\paragraph{Conformal invariance.}
A conformal field theory is invariant under conformal transformations. These are generated by conformal Killing vector fields whose infinitesimal action was briefly discussed around \eqref{eq:delta g intro}. On the complex plane equipped with the flat metric \eqref{eq:flat metric}, a holomorphic conformal transformation is generated by a coordinate transformation
\begin{equation}
\label{eq:holomorphic transformations}
z \mapsto \Pi(z)\,,
\end{equation} 
followed by a Weyl rescaling \eqref{eq:Weyl rescaling} with conformal factor
\begin{equation}
\label{eq:omega factor}
\omega=- \ln \dz \Pi(z)\,.
\end{equation}  
In this way, the background metric is indeed left invariant,
\begin{equation}
dz\, d\zbar \mapsto d\Pi\, d\zbar=\dz \Pi(z)\, dz\, d\zbar \mapsto e^\omega d\Pi\, d\zbar= dz\, d\zbar\,.
\end{equation}
The same holds for anti-holomorphic conformal transformations generated by $\zbar \mapsto \Pibar(\zbar)$. Due to the Weyl anomaly, one could have feared that conformal transformations are not symmetries of the Polyakov action, and therefore not true symmetries of the quantum theory. This is however not the case, and it can be explicitly checked that the second term in \eqref{eq:Gamma transformation} vanishes provided that $\Pi(z)$ reduces to a PSL(2,$\mathbb{C}$) global conformal transformation at infinity,
\begin{equation}
\label{eq:global conf transf}
\lim\limits_{z \to \infty}\Pi(z)=\frac{az+b}{cz+d}\,, \qquad ad-bc=1\,.
\end{equation} 

It is instructive to compute the energy in the family of vacua related by the above conformal transformations. By convention, the vacuum energy on the complex plane is normalized to zero and the corresponding value of the Liouville field therefore vanishes. Performing a conformal transformation induces a Weyl rescaling with conformal factor \eqref{eq:omega factor}. As mentioned below \eqref{eq:Weyl rescaling}, the Liouville field shifts to the new value
\begin{equation}
\phi=-\omega=\ln \partial_z \Pi(z)\,,
\end{equation} 
such that, on the manifold obtained by conformal transformation from the complex plane, the vacuum energy \eqref{eq:Tij} reduces to
\begin{equation}
\langle T(z,\zbar) \rangle_{\Pi^{-1}(\text{plane})}=\frac{c}{12}\left(\frac{\partial_z^3 \Pi}{\partial_z \Pi}-\frac{3}{2}\left(\frac{\partial_z^2 \Pi}{\partial_z \Pi}\right)^2\right)\,.
\end{equation}
We have recovered the well-known expression in terms of the Schwarzian derivative of $\Pi$.

\section{The nonlinear action governing reparametrization modes}
\label{section:effective action}
Having reviewed the Polyakov action and its basic properties, we are now ready to derive the nonlinear version of the effective action \eqref{eq:Haehl} governing reparametrization modes. Much in the same way that the \textit{infinitesimal} reparametrization mode $\epsilon^i(x)$ is defined  as parametrizing the infinitesimal metric variation \eqref{eq:delta g intro}, the \textit{finite} reparametrization mode $\Pi(z,\zbar)$ is defined as parametrizing a change of coordinate 
\begin{equation}
z \mapsto \Pi(z,\zbar)\,, 
\end{equation} 
followed by a Weyl rescaling \eqref{eq:Weyl rescaling} with parameter
\begin{equation}
\omega=- \ln \dz \Pi(z,\zbar)\,.
\end{equation} 
This yields the deformed background metric
\begin{equation}
ds^2=dz\, d\zbar+\frac{\partial_{\zbar} \Pi}{\partial_z \Pi}\, d\zbar^2\,.
\end{equation}
Of course, when $\Pi$ is a holomorphic function, it is a symmetry parameter with vanishing action. For generic configurations however, it induces a nontrivial transformation of the background metric and generates a  nonzero curvature. With the help of \eqref{eq:Gamma transformation}, we can evaluate the Polyakov action associated with this curved metric, resulting in
\begin{equation}
\label{eq:AS}
W\left[\Pi(z,\zbar)\right]=\frac{c}{48\pi} \int d^2z\, \frac{\partial_z^2 \Pi\, \partial_{\zbar} \partial_z \Pi}{(\partial_z \Pi)^2}\,.
\end{equation}
As anticipated in \cite{Cotler:2018zff} from the study of three-dimensional gravity with AdS asymptotics, this nonlinear extension of \eqref{eq:Haehl} is a complex version of the Alekseev--Shatashvili action, which was originally understood as the action of a particle on the vacuum coadjoint orbit of the Virasoro group \cite{Alekseev:1988ce}. Here, we derived it from first principles without appealing to gravity or the AdS/CFT correspondence, simply starting from the Polyakov action. Of importance for the reparametrization mode formalism to be discussed in section~\ref{section:exchanges}, up to boundary terms the action \eqref{eq:AS} is invariant under PSL(2,$\mathbb{C}$) transformations \cite{Alekseev:1988ce},
\begin{equation}
\label{eq:PSL invariance}
\Pi \mapsto \frac{a(\zbar) \Pi+b(\zbar)}{c(\zbar)\Pi+d(\zbar)}\,, \qquad ad-bc=1\,.
\end{equation}

In the remainder of this section, we illustrate how the action \eqref{eq:AS}, still viewed as a generating functional, can be used to compute stress tensor correlations on the plane or on the cylinder. 

\paragraph{Correlations on the plane.}
We recall that the reparametrization mode $\Pi$ generates a coordinate transformation followed by a Weyl rescaling. If we consider an infinitesimal version of such a transformation around the identity,
\begin{equation}
\label{eq:expansion plane}
\Pi(z,\zbar)=z+\epsilon(z,\zbar)\,,
\end{equation}
it induces the metric variation
\begin{equation}
\label{eq:source plane}
\delta g^{zz}=-4\partialbar \epsilon\,, \qquad \delta g^{\zbar\zbar}=\delta g^{z\zbar}=0\,.
\end{equation}
We should therefore expect that the action \eqref{eq:AS}, when expanded in powers of $\epsilon$, allows to compute correlations of $T_{zz}$ on the plane. Of course,  $T_{\zbar\zbar}$ correlations can be computed from $W\left[\Pibar\right]=(W\left[\Pi\right])^*$ obtained by complex conjugation. For instance, to cubic order in $\epsilon$ the action reduces to 
\begin{equation}
\label{eq:quadratic action plane}
W=-\frac{c}{48\pi} \int d^2z \left( \partial^3 \epsilon\, \partialbar \epsilon-2\, \partial^2 \epsilon\, \partial^2 \epsilon\, \partialbar \epsilon \right) +O(\epsilon^4)\,.
\end{equation}
The quadratic piece coincides with the action \eqref{eq:Haehl} when evaluated in flat complex coordinates. Treating \eqref{eq:quadratic action plane} as the generating functional with sources \eqref{eq:source plane}, one recovers the correct expressions for two- and three-point functions on the plane,\footnote{One simple way to proceed is to use the magic formula \eqref{eq:magic} in order to express \eqref{eq:quadratic action plane} as a nonlocal functional of the field $\partialbar \epsilon$ alone. For the quadratic part of the generating functional for instance, we have
\begin{equation}
\nonumber
\int d^2z_1\, \partial^3 \epsilon_1\, \partialbar \epsilon_1=\frac{1}{2\pi} \int d^2z_1 d^2z_2\, \partialbar_1 \left(\frac{1}{z_{12}}\right)  \partial^3 \epsilon_1\, \partialbar \epsilon_2=\frac{1}{2\pi} \int d^2z_1 d^2z_2\, \partial_1^3 \left(\frac{1}{z_{12}}\right)  \partialbar \epsilon_1\, \partialbar \epsilon_2\,.
\end{equation}}
\begin{align}
\label{eq:TT correlation}
\langle T(z_1,\zbar_1) T(z_2,\zbar_2) \rangle&=(-2\pi)^2\frac{\delta^2 W}{\delta \partialbar \epsilon_1\, \delta \partialbar \epsilon_2}\bigg|_{\epsilon=0}= \frac{c}{2z_{12}^4}\,,\\
\langle T(z_1,\zbar_1) T(z_2,\zbar_2) T(z_3,\zbar_3) \rangle&=(-2\pi)^3\frac{\delta^3 W}{\delta \partialbar \epsilon_1\, \delta \partialbar \epsilon_2\, \delta \partialbar \epsilon_3}\bigg|_{\epsilon=0}= \frac{c}{z_{12}^2z_{13}^2z_{23}^2}\,.
\end{align}
Thus, the Alekseev--Shatashvili action \eqref{eq:AS} is the generating functional for correlation functions of the \textit{holomorphic} stress tensor component.

\paragraph{Correlations on the cylinder.} The Alekseev--Shatashvili action can be used to compute stress tensor correlation functions on manifolds related to the plane by a conformal transformation. We illustrate this for the cylinder, covered by the real coordinates
\begin{equation}
\tau \in \mathbb{R}, \qquad \sigma \in \left[0,\beta\right)\,,
\end{equation}
As is well-known, one can map the plane to the cylinder by a conformal transformation associated with the change of coordinate 
\begin{equation}
\label{eq:Pi cylinder}
\Pi(z)=e^{-i\frac{2\pi}{\beta}z}\,, \qquad z=\sigma+i \tau\,.
\end{equation} 
Said differently, $\Pi(z)$ is now the coordinate covering the plane while $z$ is the coordinate covering the cylinder. In order to compute stress tensor correlations on the cylinder from the Alekseev--Shatashvili action \eqref{eq:AS}, we need to consider infinitesimal reparametrization modes on top of the finite conformal mapping \eqref{eq:Pi cylinder}. This is conveniently achieved by writing
\begin{equation}
\label{eq:f}
\Pi(z,\zbar)=e^{-i\frac{2\pi}{\beta}f(z,\zbar)}\,, \qquad f(z,\zbar)=z+\epsilon(z,\zbar)\,,
\end{equation}
where $\epsilon$ is periodic and asymptotes to a constant at infinity,
\begin{equation}
\label{eq:cylinder conditions}
\epsilon(\sigma+\beta,\tau)=\epsilon(\sigma,\tau)\,, \qquad \lim\limits_{\tau \to \pm \infty} \epsilon(\sigma,\tau)=\text{cst}\,.
\end{equation}
Plugging \eqref{eq:f} into the nonlinear action \eqref{eq:AS} and making use of the conditions \eqref{eq:cylinder conditions} to discard total derivative terms, we obtain
\begin{equation}
\label{eq:AS cylinder}
W\left[f(z,\zbar)\right]=\frac{c}{48\pi} \int d^2z \left(-\left(\frac{2\pi}{\beta}\right)^2 \partial_z f\, \partial_{\zbar} f+ \frac{\partial_z^2 f\, \partial_{\zbar} \partial_z f}{(\partial_z f)^2}\right)\,.
\end{equation}
Interestingly, this alternative form of the Alekseev--Shatashvili action coincides with the action of a particle on the first exceptional coadjoint orbit of the Virasoro group \cite{Alekseev:1988ce}.\footnote{I thank Jordan Cotler and Jakob Salzer for discussions on this point.} Note that it naturally inherits the PSL(2,$\mathbb{C}$) invariance described in \eqref{eq:PSL invariance} through the identification $\Pi=e^{-i\frac{2\pi}{\beta} f}$. As for the plane, we expand this action in powers of the infinitesimal reparametrization mode $\epsilon$ \eqref{eq:f} that is appropriate to the cylinder. To quadratic order, we get
\begin{equation}
W=-\frac{c}{48\pi} \left(\frac{2\pi}{\beta}\right)^2 \int d^2z \left(\partialbar \epsilon+\partial \epsilon\,  \partialbar \epsilon+\left(\frac{\beta}{2\pi}\right)^2\partial^3 \epsilon\, \partialbar \epsilon \right)+O(\epsilon^3)\,.
\end{equation}
In particular, we recover the quadratic action for reparametrizations of the thermal cylinder constructed in \cite{Haehl:2019eae}. In addition, we also find a linear term which turns out to account for the Casimir energy of the cylinder,
\begin{align}
\langle T(z,\zbar) \rangle =-2\pi\, \frac{\delta W}{\delta \partialbar \epsilon(z,\zbar)}\bigg|_{\epsilon=0}=\frac{c}{24} \left(\frac{2\pi}{\beta}\right)^2\,.
\end{align}
We can similarly recover the stress tensor two-point function,
\begin{subequations}
\begin{align}
\langle T(z_1,\zbar_1) T(z_2,\zbar_2) \rangle&=(2\pi)^2 \frac{\delta^2 W}{\delta \partialbar \epsilon_1\, \delta \partialbar \epsilon_2}\bigg|_{\epsilon=0}=-\frac{c}{12} \left[\left(\frac{2\pi}{\beta}\right)^2 \partial_1 +\partial_1^3\right] \left(\frac{1}{z_{{12}}}\right)\\
&=\frac{c}{2} \left[\frac{1}{z_{12}^4}+\left(\frac{2\pi}{\beta}\right)^2 \frac{1}{6z_{12}^2}\right]=\left(\frac{\pi}{\beta}\right)^4 \frac{c}{2\sin^4 \frac{\pi}{\beta}z_{12}}\,,
\end{align}
\end{subequations}
where the last equality holds up to non-singular terms that are irrelevant. 

In summary, in this section we have derived the nonlinear extension of the generating functional \eqref{eq:Haehl}, starting from the Polyakov action. We have illustrated how it can be used to compute correlations of the holomorphic stress tensor component on manifolds related to the complex plane by a conformal transformation. 

\section{The effective theory of stress tensor exchanges}
\label{section:exchanges}
We now come to the description of stress tensor exchanges in the reparametrization mode formalism, arguably its main interest from a computational perspective. As usual, the holomorphic and anti-holomorphic dependencies of correlation functions factorize. For simplicity, we will only discuss their holomorphic part, but an analogous reasoning obviously applies to their anti-holomorphic counterpart as well. 

We start by reviewing the prescriptions for computing Virasoro identity blocks within the reparametrization formalism, following \cite{Haehl:2019eae,Cotler:2018zff,Anous:2020vtw}. Ultimately, our goal will be to \textit{derive} these rules from Feynman diagrams describing stress tensor exchanges between external primary operators. As a preliminary step to the reparametrization formalism, one considers primary two-point functions on a manifold related to the complex plane by a conformal transformation \eqref{eq:holomorphic transformations} with symmetry parameter $\Pi(z)$,
\begin{equation}
\label{eq:reparametrized 2point}
\langle \Oh(1) \Oh(2) \rangle_{\Pi^{-1}(\text{plane})} =\left(\frac{\partial_{z_1} \Pi(z_1)\, \partial_{z_2} \Pi(z_2)}{\left(\Pi(z_1)-\Pi(z_2)\right)^2}\right)^{h}\,.
\end{equation}
A \textit{bilocal vertex operator} $\Bh$\footnote{It has been recently argued that the bilocal operator \eqref{eq:bilocal vertex} can be formally identified with the Virasoro identity OPE block \cite{DHoker:2019clx}. It is formal in the sense that it still requires proper renormalization.} is then introduced by promoting the symmetry parameter $\Pi(z)$ to an arbitrary reparametrization mode $\Pi(z,\zbar)$,
\begin{equation}
\label{eq:bilocal vertex}
\Bh(1,2)\equiv \left(\frac{\partial_{z_1} \Pi(z_1,\zbar_1)\, \partial_{z_2} \Pi(z_2,\zbar_2)}{\left(\Pi(z_1,\zbar_1)-\Pi(z_2,\zbar_2)\right)^2}\right)^{h}\,.
\end{equation}
For the purpose of computing stress tensor exchanges on the plane, we again expand $\Pi(z,\zbar)$ around the identity. As was pointed out in \cite{Anous:2020vtw}, an infinitesimal reparametrization mode $\epsilon$ exponentiates into a finite mode through $\Pi=e^{\epsilon \partial} z=z+\epsilon+\frac{1}{2}\epsilon \partial \epsilon+...$ such that the expansion of the bilocal vertex \eqref{eq:bilocal vertex} takes the form
\begin{equation}
\label{eq:vertex expansion}
\Bh(1,2)= \frac{1}{(z_{12})^{2h}} \sum_{n\geq 0} \Bh^{(n)}(1,2)\,,
\end{equation}
where the first few terms are given by
\begin{subequations}
\begin{align}
\Bh^{(0)}(1,2)&=1\,,\\
\Bh^{(1)}(1,2)&=b^{(1)}_h(1,2)\,,\\
\Bh^{(2)}(1,2)&=\frac{1}{2!}\left(b^{(1)}_h(1,2)\right)^2+b^{(2)}_h(1,2)\,,\\
\nonumber &\hskip 2mm \vdots\\
\Bh^{(n)}(1,2)&=\frac{1}{n!}\left(b^{(1)}_h(1,2)\right)^n + \text{lower orders in } h\,,
\end{align}
\end{subequations}
with
\begin{subequations}
\label{eq:b vertices}
\begin{align}
b^{(1)}_h(1,2)&=h\left(\partial \epsilon_1+\partial \epsilon_2-2\frac{\epsilon_1-\epsilon_2}{z_{12}}\right)\,,\\
b^{(2)}_h(1,2)&=h\left(\frac{\epsilon_1\partial^2 \epsilon_1+\epsilon_2\partial^2 \epsilon_2}{2}-\frac{\epsilon_1\partial \epsilon_1-\epsilon_2\partial \epsilon_2}{z_{12}}+\frac{(\epsilon_1-\epsilon_2)^2}{z_{12}^2}\right)\,.
\end{align}
\end{subequations}
More details and higher order terms of this expansion are found in \cite{Anous:2020vtw}. 

Remarkably, the bilocal vertex operator $\Bh$ can be used to straightforwardly compute the contribution of the Virasoro identity block $\mathcal{V}_0$ to four-point functions involving pairs of identical operators,
\begin{equation}
\label{eq:Virasoro blocks}
\langle V(1) V(2) W(3) W(4) \rangle=\frac{1}{(z_{12})^{2h_V} (z_{34})^{2h_W}} \sum_{\mathcal{O}} C_{VV\mathcal{O}}\, C_{WW\mathcal{O}}\,  \mathcal{V}_{h_{\mathcal{O}}}(u)\,,
\end{equation}
where $u= \frac{z_{12}z_{34}}{z_{13}z_{24}}$ and we again only displayed the holomorphic part of the correlator. In \eqref{eq:Virasoro blocks}, the four-point function is expressed as a sum over Virasoro conformal blocks $\mathcal{V}_{h_\mathcal{O}}(u)$, where the sum runs over all primary operators $\mathcal{O}$ and where $C_{VV\mathcal{O}}$ and $C_{WW\mathcal{O}}$ are fusion coefficients that characterize any particular CFT. Note that conformal blocks are purely kinematical objects that only depend on the conformal dimensions $h_V, h_W, h_{\mathcal{O}}$ of the various operators involved. A prescription to compute the contribution from the Virasoro identity block $\mathcal{V}_0$ based on the reparametrization mode formalism has been put forward in \cite{Haehl:2018izb,Cotler:2018zff,Haehl:2019eae} and further developed in \cite{Anous:2020vtw}. It involves the following ingredients:
\begin{itemize}
	\item A re-interpreation of the reparametrization mode $\epsilon(z,\zbar)$ as a \textit{dynamical} field instead of as a source for the stress tensor, together with a re-interpretation of the action \eqref{eq:quadratic action plane} as that governing its dynamics. Accordingly, the reparametrization mode propagator is found to be
	\begin{equation}
	\label{eq:unphysical prop}
	\langle \epsilon(z_1,\zbar_1) \epsilon(z_2,\zbar_2) \rangle=\frac{6}{c} z_{12}^2 \ln \mu |z_{12}|\,,
	\end{equation}
	with $\mu$ an arbitrary energy scale that cannot be determined from the theory but eventually drops out from the four-point functions of interest.
	
	\item A \textit{gauging} of the PSL(2,$\mathbb{C}$) symmetry \eqref{eq:PSL invariance}, resulting in the physical gauge-invariant and purely holomorphic propagator 
	\begin{equation}
	\label{eq:physical prop}
	G_\epsilon(z_1,z_2)\equiv \langle \epsilon(z_1,\zbar_1) \epsilon(z_2,\zbar_2) \rangle_{\text{phys}}=\frac{6}{c} z_{12}^2 \ln \mu z_{12}\,.
	\end{equation}
	This physical propagator can be alternatively obtained by a monodromy projection of \eqref{eq:unphysical prop} as described in \cite{Haehl:2019eae}.
	
	\item The identification of the Virasoro identity block as a \textit{connected} correlation function of bilocal vertex operators,
	\begin{equation}
	\label{eq:rule 3}
	\frac{C_{VVT}\, C_{WWT}}{(z_{12})^{2h_V} (z_{34})^{2h_W}}\, \mathcal{V}_0(u)\equiv \langle \mathcal{B}_{h_V}(1,2) \mathcal{B}_{h_W}(3,4) \rangle_c\,,
	\end{equation}
	where the vertices are viewed as functionals of the dynamical reparametrization field $\epsilon$ with physical propagator \eqref{eq:physical prop}. The fact that this universal formula describes the fusion coefficients $C_{VVT}, C_{WWT}$ together with the Virasoro identity block $\mathcal{V}_0$ should not come as a surprise, since the coupling between primary operators and the stress tensor is universally dictated by conformal symmetry.
\end{itemize} 

Following the above set of prescribed rules, the Virasoro identity block contribution to the \textit{normalized} four-point function
\begin{equation}
\label{eq:F4}
\mathcal{F}_4\equiv \frac{\langle V(1) V(2) W(3) W(4) \rangle}{\langle V(1) V(2) \rangle \langle W(3) W(4) \rangle}\bigg|_{\mathcal{V}_0}\,,
\end{equation}
can be computed order by order in a $1/c$ expansion. Indeed, each propagator $G_\epsilon$ comes with a factor of $1/c$ such that the reparametrization formalism naturally organizes as a perturbative expansion at large central charge. In addition, terms in the expansion of the bilocal vertex operator \eqref{eq:vertex expansion} that contribute to a given order in $1/c$ are easy to identify. At zeroth order, we simply have
\begin{equation}
\mathcal{F}_4\big|_{O(1)}=\langle \mathcal{B}_{h_V}^{(0)}(1,2) \mathcal{B}_{h_W}^{(0)}(3,4) \rangle =1\,.
\end{equation}
At subleading order, we have \cite{Cotler:2018zff}
\begin{subequations}
\label{eq:F4 subleading}
\begin{align}
\mathcal{F}_4\big|_{O(1/c)}&=\langle \mathcal{B}_{h_V}^{(1)}(1,2) \mathcal{B}_{h_W}^{(1)}(3,4) \rangle=\langle b^{(1)}_{h_V}(1,2) b^{(1)}_{h_W}(3,4) \rangle\\
&=\frac{2 h_V h_W}{c}\, u^2\, \, {}_2F_1\left(2,2,4;u\right)\,,
\end{align}
\end{subequations}
which is recognized as the \textit{global} identity block contribution \cite{Dolan:2000ut,Dolan:2003hv}. We will also discuss the terms appearing at order $O(1/c^2)$ without explicitly evaluating them, 
\begin{align}
\label{eq:F4 subsubleading}
\mathcal{F}_4\big|_{O(1/c^2)}&=\langle \mathcal{B}_{h_V}^{(2)}(1,2) \mathcal{B}_{h_W}^{(2)}(3,4) + \mathcal{B}_{h_V}^{(3)}(1,2) \mathcal{B}_{h_W}^{(1)}(3,4) + \mathcal{B}_{h_V}^{(1)}(1,2) \mathcal{B}_{h_W}^{(3)}(3,4) \rangle_c\\
\nonumber
&+\langle \mathcal{B}^{(2)}_{h_V}(1,2) \mathcal{B}^{(1)}_{h_W}(3,4) + \mathcal{B}^{(1)}_{h_V}(1,2) \mathcal{B}^{(2)}_{h_W}(3,4) \rangle_c \,.
\end{align}
Upon replacement of the bilocal vertices by their expressions in terms of reparametrization modes, the first line contains terms involving two propagators $\langle \epsilon \epsilon \rangle$ while the second line contains terms involving a single three-point function $\langle \epsilon \epsilon \epsilon \rangle$\footnote{The three-point function $\langle \epsilon \epsilon \epsilon \rangle$, whose explicit expression may be found in \cite{Anous:2020vtw}, scales like $1/c^2$.}. Except from the first one, all terms suffer from ultraviolet (UV) divergences since they contain $\epsilon$ correlators evaluated at coincident points. To make sense of these, one would need to supplement the reparametrization formalism with a regularization procedure. We will not try to remedy this here, and we will restrict our attention to the first regular term instead. We make a few comments regarding divergences and their regularization in the discussion section.  

As shown in \cite{Cotler:2018zff}, the above prescription successfully reproduces known results at large central charge $c$ in the `light-light' limit $h_V, h_W =O(\sqrt{c})$ \cite{Fitzpatrick:2014vua} and in the `heavy-light' limit $h_V=O(1),\, h_W=O(c)$ \cite{Fitzpatrick:2015zha,Fitzpatrick:2015dlt,Beccaria:2015shq}. Furthermore, an appropriate modification of the reparametrization mode formalism to Lorentzian signature similarly led to a successful description of the maximal Lyapunov growth displayed by out-of-time-order (OTOC) correlators at large central charge \cite{Haehl:2018izb,Cotler:2018zff,Haehl:2019eae}.

In spite of these successes, the origin of the above set of rules seems rather mysterious at first sight. A convincing justification was nonetheless provided by Haehl, Reeves and Rozali by showing that these rules are those of the \textit{shadow operator formalism} upon identification of $\epsilon(z,\zbar)$ with the shadow of the stress tensor $T(z,\zbar)$ \cite{Haehl:2019eae}. However, their argument only applied to the first nontrivial term in the bilocal vertex expansion \eqref{eq:vertex expansion}, i.e., to $\Bh^{(1)}(1,2)$. Hence, this argument guarantees that the global identity block \eqref{eq:F4 subleading} is correctly accounted for, but a justification of the validity of the reparametrization mode formalism at all orders in perturbation is still missing.

We wish to provide a derivation of the above set of rules which can be extended to higher perturbative orders, and which does not refer to the shadow operator formalism at any step. In particular, we shall not need to re-interpret $\epsilon(z,\zbar)$ as a dynamical field, which we find somewhat awkward given its meaning of source for the stress tensor when first introduced. The alternative derivation which we propose is straightforward and simply consists in computing contributions to the normalized four-point function \eqref{eq:F4} from (position-space) Feynman diagrams involving stress tensor exchanges between the two pairs of identical operators. These diagrams are shown in Figure~\ref{fig:Feynman}. As will be shown below, we find perfect agreement with the reparametrization formalism. We believe that the alternative method developed below conceptually clarifies the results of the reparametrization mode formalism, and provides compelling evidence of its validity. 

\begin{figure}
	\centering
	\begin{subfigure}[b]{.49\textwidth}
		\centering
		\includegraphics[scale=0.4]{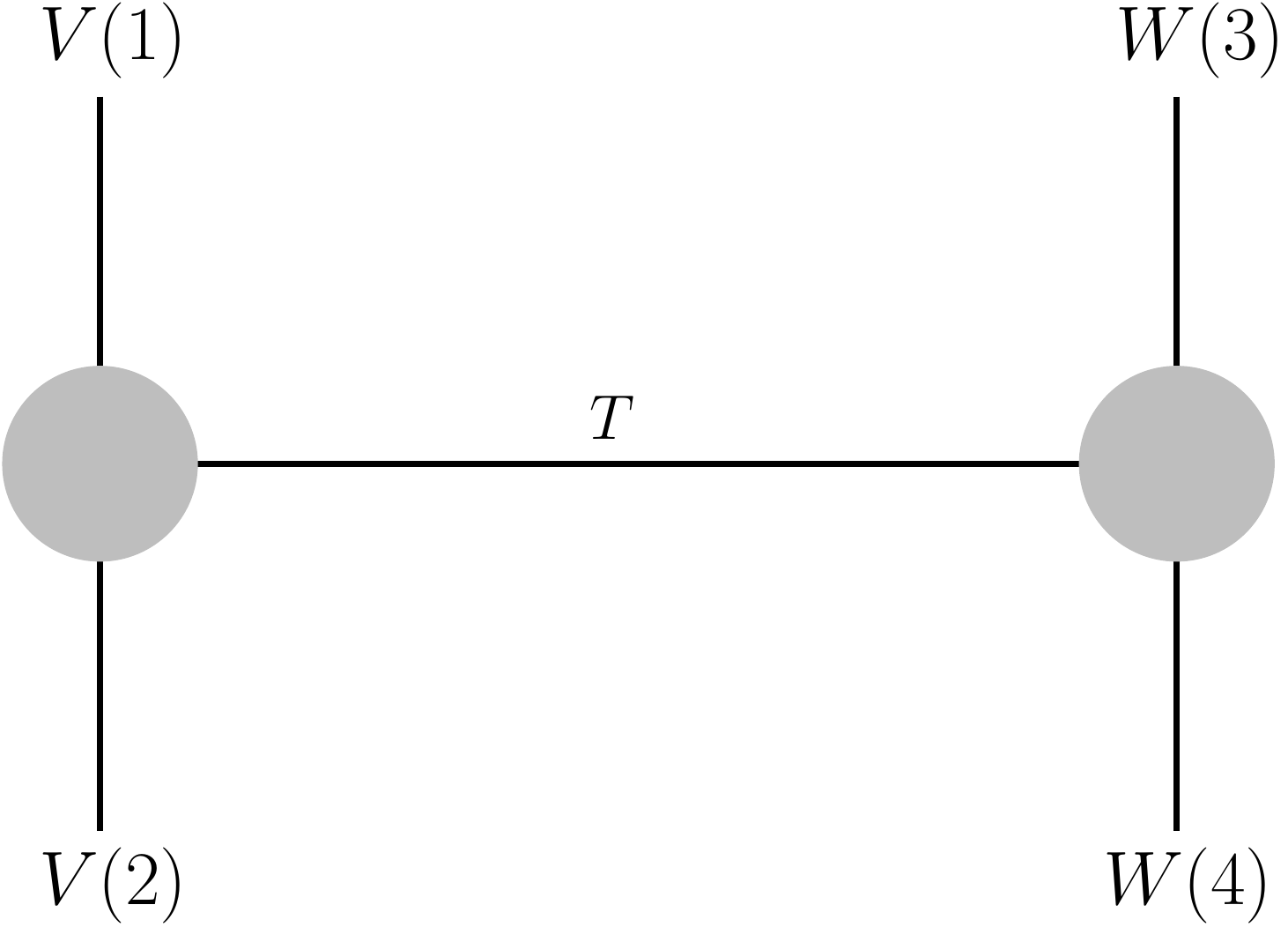}
		\caption{}
		\label{fig:Feynman1}
	\end{subfigure}
	\begin{subfigure}[b]{.49\textwidth}
		\centering
		\includegraphics[scale=0.4]{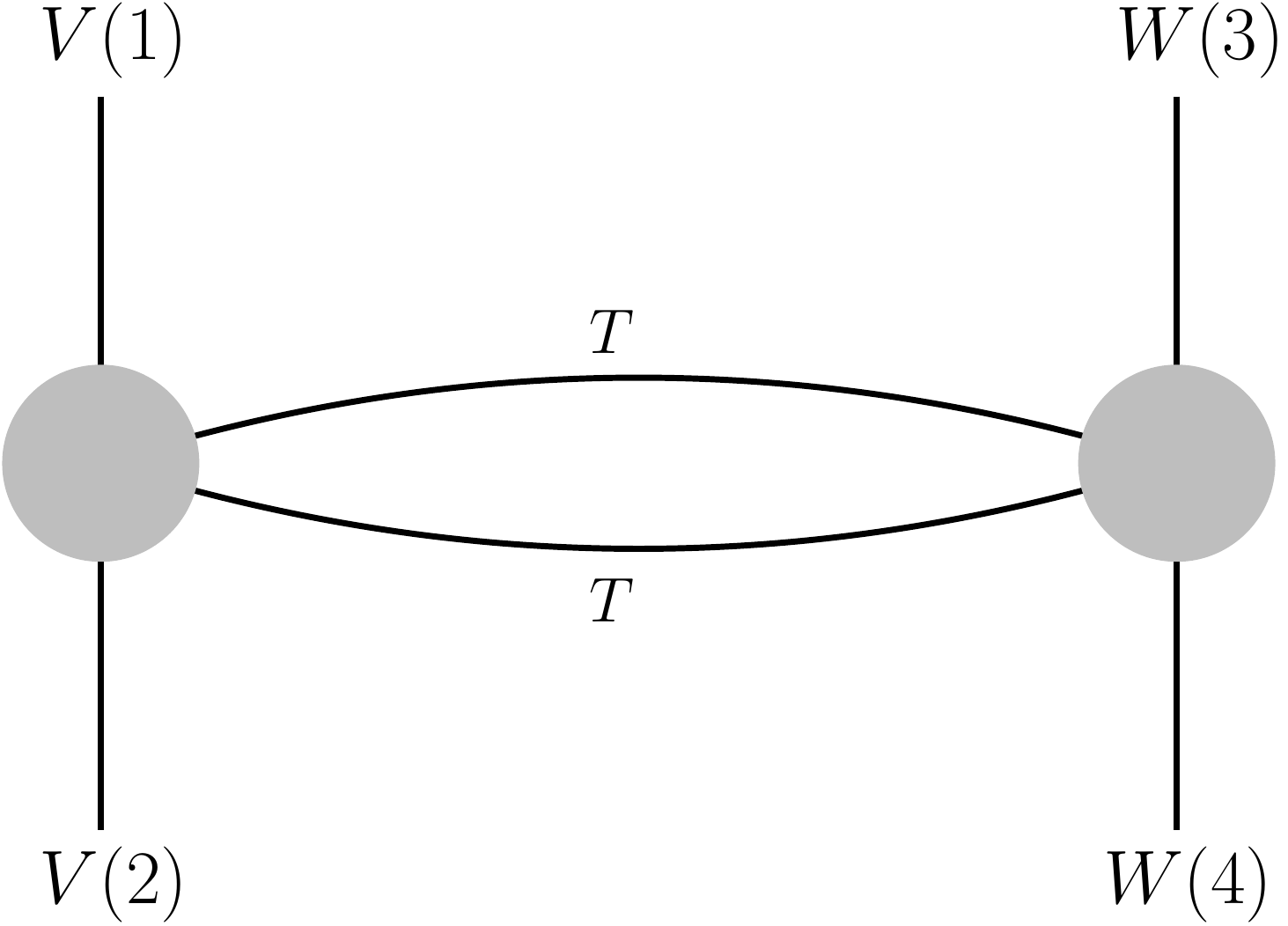}
		\caption{}
		\label{fig:Feynman2}
	\end{subfigure}\\
	\begin{subfigure}[b]{.49\textwidth}
		\centering
		\includegraphics[scale=0.4]{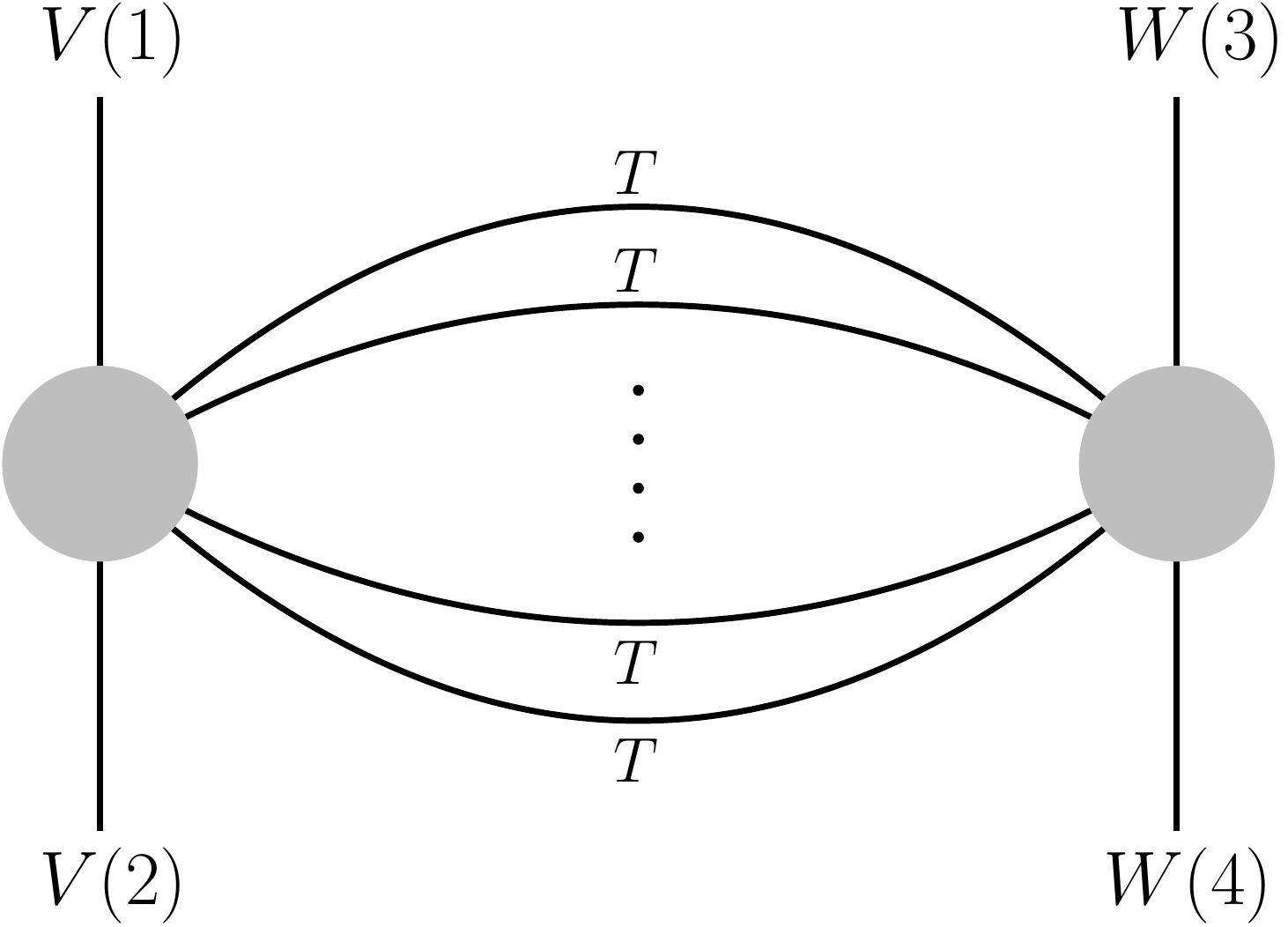}
		\caption{}
		\label{fig:Feynman3}
	\end{subfigure}\\
	\caption{Feynman diagrams corresponding to stress tensor exchanges between two pairs of identical operators. The large grey circles refer to the vertices $\langle V(1)V(2) \hat{T}(w_1)... \hat{T}(w_n) \rangle$, which are \textit{nonlocal} and \textit{exact} (as opposed to free vertices). (a) Single exchange diagram~$\mathcal{A}_1$. (b) Double exchange diagram $\mathcal{A}_2$. (c) Multiple exchange diagram $\mathcal{A}_n$.}
	\label{fig:Feynman}
\end{figure}

The atomic ingredients that we need are the stress tensor propagator \eqref{eq:TT correlation} together with the \textit{partially amputated} $(2+n)$-point correlation functions
\begin{gather}
\label{eq:n+2 point}
\langle V(1) V(2) \hat{T}(w_1)\, ...\, \hat{T}(w_n) \rangle=\left(-\frac{6}{\pi c}\right)^n\, \prod_{i=1}^{n} \partial_{\wbar_i} (\partial_{w_i})^{-3}\, \langle V(1) V(2) T(w_1) ... T(w_n) \rangle\,,
\end{gather}
where the two matter insertions on the left-hand side are unamputated while the $n$ stress tensor insertions are amputated. We denote amputated stress tensor insertions with a hat. The above equality can be derived by first writing a correlator with unamputated $i$-th leg as the convolution of its amputated counterpart with the stress tensor propagator, 
\begin{align}
\langle ...\, T(w_i)\, ... \rangle &=\int d^2y\, \langle ...\, \hat{T}(y)\, ... \rangle  \langle T(y) T(w_i) \rangle\,.
\end{align}
Equation \eqref{eq:n+2 point} is obtained after invoking the magic identity \eqref{eq:magic} in order to rewrite the stress tensor propagator as
\begin{equation}
\label{eq:stress tensor prop}
\langle T(z) T(w) \rangle=\frac{c}{2(z-w)^4}=-\frac{c}{12}\, \partial_w^3 \left(\frac{1}{z-w}\right)=-\frac{\pi c}{6}\, \partial_w^3\, (\partial_{\bar{w}})^{-1} \delta(z-w)\,.
\end{equation}
The partially amputated correlation functions \eqref{eq:n+2 point} will be used as vertices in evaluating the Feynman diagrams of interest. Because they are \textit{exact} rather than free vertices, we indicate them with large grey circles in Figure~\ref{fig:Feynman}.  

Before turning to their evaluation, let us comment on the overall power of $1/c$ associated with a Feynman diagram involving $n$ stress tensor exchanges. Such a diagram contains two vertices \eqref{eq:n+2 point} and $n$ stress tensor propagators \eqref{eq:stress tensor prop}, so that it has an overall factor of $(1/c)^n$. In the reparametrization mode formalism, this factor would be associated to $n$ reparametrization propagators \eqref{eq:physical prop}. Of course, we will discover that this is not a coincidence.

\paragraph{Single exchange.} 
We first evaluate the Feynman diagram of Figure~\ref{fig:Feynman1} containing a single stress tensor exchange. Patching together the vertices \eqref{eq:n+2 point} and the stress tensor propagator, we have
\begin{gather}
\label{eq:single}
\mathcal{A}_1= \int d^2w_1\, d^2w_2\, \langle V(1)V(2) \hat{T}(w_1) \rangle \langle T(w_1) T(w_2) \rangle \langle \hat{T}(w_2) W(3) W(4) \rangle\\
\nonumber
= \left(\frac{6}{\pi c}\right)^2 \int d^2w_1\, d^2w_2\, \langle V(1)V(2) \partialbar T(w_1) \rangle \partial_{w_1}^{-3} \partial_{w_2}^{-3} \langle T(w_1) T(w_2) \rangle \langle \partialbar T(w_2) W(3) W(4) \rangle\,.
\end{gather}
Remarkably, we observe that the kernel of the second line coincides with the `physical reparametrization propagator' \eqref{eq:physical prop},
\begin{equation}
\partial_{w_1}^{-3} \partial_{w_2}^{-3} \langle T(w_1) T(w_2) \rangle=\frac{c}{24} w_{12}^2 \ln \mu w_{12}=\left(\frac{c}{12}\right)^2 G_\epsilon(w_1,w_2)\,.
\end{equation}
We stress that we never had to consider any kind of coupling to a dynamical reparametrization mode in order to witness the appearance of this propagator. In this approach, $G_\epsilon$ is an `emergent' quantity derived from the stress tensor propagator.  
To simplify \eqref{eq:single} further, we use the conformal Ward identity
\begin{subequations}
\label{eq:conformal Ward}
\begin{align}
\langle \partialbar T(w) V(1) V(2) \rangle&=-2\pi \sum_{i=1,2} \left[h_V\, \partial_w \delta^{(2)}(w-z_i)-\delta^{(2)}(w-z_i) \partial_{z_i}\right] \langle V(1) V(2) \rangle\\
&=-2\pi h_V \langle V(1) V(2) \rangle \left[\left(\partial_w +\frac{2}{z_{12}}\right) \delta(w-z_1)+(z_1 \leftrightarrow z_2)\right]\,.
\end{align}
\end{subequations}
Plugging \eqref{eq:conformal Ward} into \eqref{eq:single} and integrating by parts, we find
\begin{align}
\label{eq:single 2}
\mathcal{F}_4\big|_{O(1/c)}=\frac{\mathcal{A}_1}{\langle VV \rangle \langle WW \rangle}=\int d^2w_1\, d^2w_2\, D^{h_V}_{w_1}(1,2) D^{h_W}_{w_2}(3,4)\,  G_\epsilon(w_1,w_2)\,,
\end{align}
where we defined the differential operator
\begin{equation}
\label{eq:D operator}
D_w^{h}(1,2)\equiv h\left[\delta(w-z_1) \left(\partial_{w}-\frac{2}{z_{12}}\right)+(z_1 \leftrightarrow z_2)\right]\,.
\end{equation}
Of course, the delta distribution in \eqref{eq:D operator} allows to trivially perform the integrals in \eqref{eq:single 2}. Doing so and comparing with the bilocal vertex operators in \eqref{eq:b vertices}, the formula \eqref{eq:F4 subleading} obtained from the reparametrization formalism emerges before our eyes,
\begin{equation}
\mathcal{F}_4\big|_{O(1/c)}=\langle b^{(1)}_{h_V}(1,2) b^{(1)}_{h_W}(3,4) \rangle\,.
\end{equation}
A successful derivation of the reparametrization mode prescription is thus provided by evaluating the Feynman diagram containing a single stress tensor exchange.

\paragraph{Double exchange.} To further test the correspondence with the reparametrization formalism uncovered at order $O(1/c)$, we evaluate the contribution coming from two stress tensor exchanges. The corresponding Feynman diagram is shown in Figure~\ref{fig:Feynman2}. Dividing by the appropriate symmetry factor of $2$ associated with the interchange of internal lines, and applying the same line of reasoning as above, we find
\begin{align}
\label{eq:F4 double}
\mathcal{A}_2=\frac{1}{2(2\pi)^4}
\int \prod_{i=1}^4 d^2w_i\,& \langle V(1)V(2) \partialbar T(w_1) \partialbar T(w_2) \rangle G_\epsilon(w_1,w_3)\\
\nonumber
&\times\,   G_\epsilon(w_2,w_4) \langle \partialbar T(w_3) \partialbar T(w_4) W(3) W(4) \rangle\,.
\end{align}
The four-point function $\langle \partialbar T \partialbar T VV\rangle$ is derived in appendix~\ref{appendix:4 point} from the conformal Ward identity. Plugging its expression in \eqref{eq:F4 double}, a tedious but straightforward computation yields
\begin{subequations}
\label{eq:F4 double 2}
\begin{align}
\nonumber
&\mathcal{F}_4\big|_{O(1/c^2)} \supset \frac{\mathcal{A}_2}{\langle VV \rangle \langle WW \rangle}\\
&=\frac{1}{2} \int \prod_{i=1}^4 d^2w_i\, D^{h_V}_{w_1}(1,2) D^{h_W}_{w_3}(3,4) G_\epsilon(w_1,w_3)\, D^{h_V}_{w_2}(1,2) D^{h_W}_{w_4}(3,4) G_\epsilon(w_2,w_4)\\
&+\int \prod_{i=1}^4 d^2w_i\, D^{h_V}_{\lbrace w_1,w_2 \rbrace}(1,2)\, D^{h_W}_{w_3}(3,4) G_\epsilon(w_1,w_3)\, D^{h_W}_{w_4}(3,4) G_\epsilon(w_2,w_4)\\
&+\int \prod_{i=1}^4 d^2w_i\, D^{h_W}_{\lbrace w_3,w_4 \rbrace}(3,4)\, D^{h_V}_{w_1}(1,2) G_\epsilon(w_1,w_3)\, D^{h_V}_{w_2}(1,2) G_\epsilon(w_2,w_4)\\
&+2\int \prod_{i=1}^4 d^2w_i\, D^{h_V}_{\lbrace w_1,w_2 \rbrace}(1,2)D^{h_W}_{\lbrace w_3,w_4 \rbrace}(3,4)\, G_\epsilon(w_1,w_3) G_\epsilon(w_2,w_4)\,,
\end{align}
\end{subequations}
where $D^h_w$ was given in \eqref{eq:D operator} and we have introduced a second differential operator,
\begin{align}
D^{h}_{w_1,w_2}(1,2)\equiv h  \bigg[\delta(w_2-z_1&)\delta(w_1-z_1)\left(\frac{\partial_{w_1}^2}{2}-\frac{\partial_{w_1}}{z_{12}}+\frac{1}{z_{12}^2}\right)\\
\nonumber
&-\frac{\delta(w_2-z_2)\delta(w_1-z_1)}{z_{12}^2}+(z_1\leftrightarrow z_2)\bigg]\,.
\end{align}
As before, the integrals in \eqref{eq:F4 double 2} localize due to the delta distributions. Like at order $O(1/c)$, we want to make the comparison with the reparametrization formalism. More precisely, because \eqref{eq:F4 double 2} involves two propagators $G_\epsilon$ whose legs are connected to \textit{both} pairs of operators, it should be compared to
\begin{align}
\label{eq:b2b2}
\langle \mathcal{B}_{h_V}^{(2)}(1,2) \mathcal{B}_{h_W}^{(2)}(3,4) \rangle_c
&=\frac{1}{2}\left(\langle b_{h_V}^{(1)}(1,2) b_{h_W}^{(1)}(3,4)\rangle \right)^2+\frac{1}{2}\langle \left(b^{(1)}_{h_W}(3,4)\right)^2\, b^{(2)}_{h_V}(1,2) \rangle_c\\
\nonumber
&+\frac{1}{2}\langle \left(b^{(1)}_{h_V}(1,2)\right)^2\, b^{(2)}_{h_W}(3,4)\rangle_c+ \langle b^{(2)}_{h_V}(1,2) b^{(2)}_{h_W}(3,4) \rangle_c\,.
\end{align} 
Looking again at the definitions of the bilocal vertices in \eqref{eq:b vertices}, a careful comparison shows that the four different terms in \eqref{eq:b2b2} exactly coincide with the four terms in \eqref{eq:F4 double 2}. The agreement occurs term by term such that \eqref{eq:F4 double 2} and \eqref{eq:b2b2} are just two ways of writing the same quantities. Once again, the reparametrization formalism has effectively emerged when evaluating Feynman diagrams describing stress tensor exchanges.

\paragraph{Exponentiation in the light-light limit.} One of the successes of the reparametrization formalism was to correctly reproduce the leading term of the Virasoro identity block in the light-light limit $h=O(\sqrt{c})$ \cite{Cotler:2018zff},
\begin{equation}
\label{eq:exponential}
\mathcal{F}_4=\exp \left(\frac{2h_V h_W}{c}\, u^2 \, {}_2F_1\left(2,2,4;u\right)\right)+O(1/\sqrt{c})\,.
\end{equation}
Hence, the Virasoro identity block contains a contribution which is the exponentiated global identity block \eqref{eq:F4 subleading}. This result was first derived in \cite{Fitzpatrick:2014vua}.

Since it is an important result, it is worth deriving it within the alternative formalism proposed here. For this, we consider the contributions resulting from an arbitrary number $n$ of stress tensor exchanges. The corresponding Feynman diagram is shown in Figure~\ref{fig:Feynman3}. We use the expression for the vertex $\langle \partialbar T(w_1)...\partialbar T(w_n) V(1)V(2) \rangle$ given in \eqref{eq:TTTVV appendix} which holds in the light-light limit $h=O(\sqrt{c})$, and integrate it against $n$ stress tensor propagators. Taking into account the symmetry factor of $n!$ associated with interchanges of internal lines, and after a straightforward computation similar to that for single and double exchanges, we find  
\begin{equation}
\mathcal{F}_4=\sum_n \frac{1}{n!} \left( \langle b^{(1)}_{h_V}(1,2) b^{(1)}_{h_W}(3,4) \rangle \right)^n+O(1/\sqrt{c})\,.
\end{equation}
Upon insertion of \eqref{eq:F4 subleading}, we indeed recover the expected exponential \eqref{eq:exponential}.

\section{Discussion}
We have discussed several aspects of the reparametrization mode formalism. After reviewing some of the basic properties of the Polyakov action in section~\ref{section:Polyakov}, we provided a first principle derivation of the Alekseev--Shatashvili action as a nonlinear extension of the effective action \eqref{eq:Haehl} governing the reparametrization modes. We have further argued in section~\ref{section:effective action} that the correct interpretation of the reparametrization mode is that of a source for the holomorphic component of the stress tensor, and that the Alekseev--Shatashvili action is the generating functional for its connected correlation functions on manifolds related to the complex plane by conformal transformations. We then turned to the computation of Virasoro identity blocks within the reparametrization mode formalism in section~\ref{section:exchanges} where we showed that the otherwise mysterious prescriptions of that formalism naturally emerge when evaluating Feynman diagrams associated with stress tensor exchanges between pairs of identical primary operators. Several interesting open problems deserve further investigation, which will help bring the program initiated here to further completion.

\paragraph{Comparison with other formalisms.} 
Although the approach proposed here and based on the evaluation of Feynman diagrams will look familiar to anyone having studied perturbative quantum field theory, it is quite unconventional from the common perspective on 2d CFTs. In fact, we are not aware of any other similar use of Feynman diagrams made in this context. It would therefore be very interesting to connect it to more conventional techniques used to compute Virasoro blocks in 2d CFTs \cite{Fitzpatrick:2014vua,Fitzpatrick:2015dlt,Chen:2016cms,Fitzpatrick:2016mtp}. In particular, the formalism developed in \cite{Fitzpatrick:2016mtp,DHoker:2019clx} based on gravitational Wilson lines and the AdS/CFT correspondence seems very close in spirit to our approach. Indeed, it was shown that the expectation value of the gravitational Wilson line coincides with the reparametrized two-point function \eqref{eq:reparametrized 2point}, while a natural interpretation in terms of Feynman diagrams and stress tensor exchanges also emerged in that picture. 

\begin{figure}
	\centering
	\begin{subfigure}[b]{.49\textwidth}
		\centering
		\includegraphics[scale=0.4]{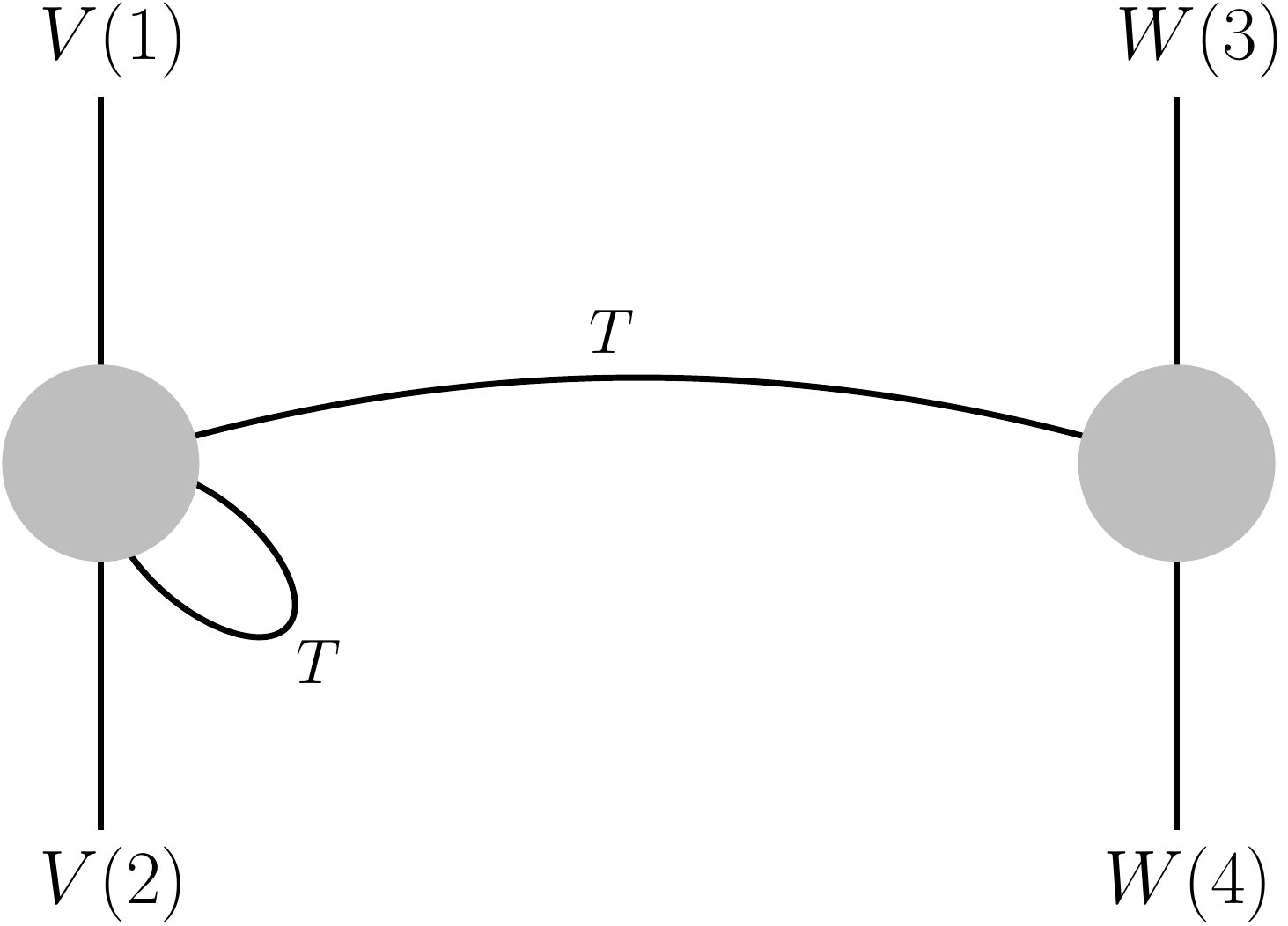}
		\caption{}
		\label{fig:B3B1}
	\end{subfigure}
	\begin{subfigure}[b]{.49\textwidth}
		\centering
		\includegraphics[scale=0.4]{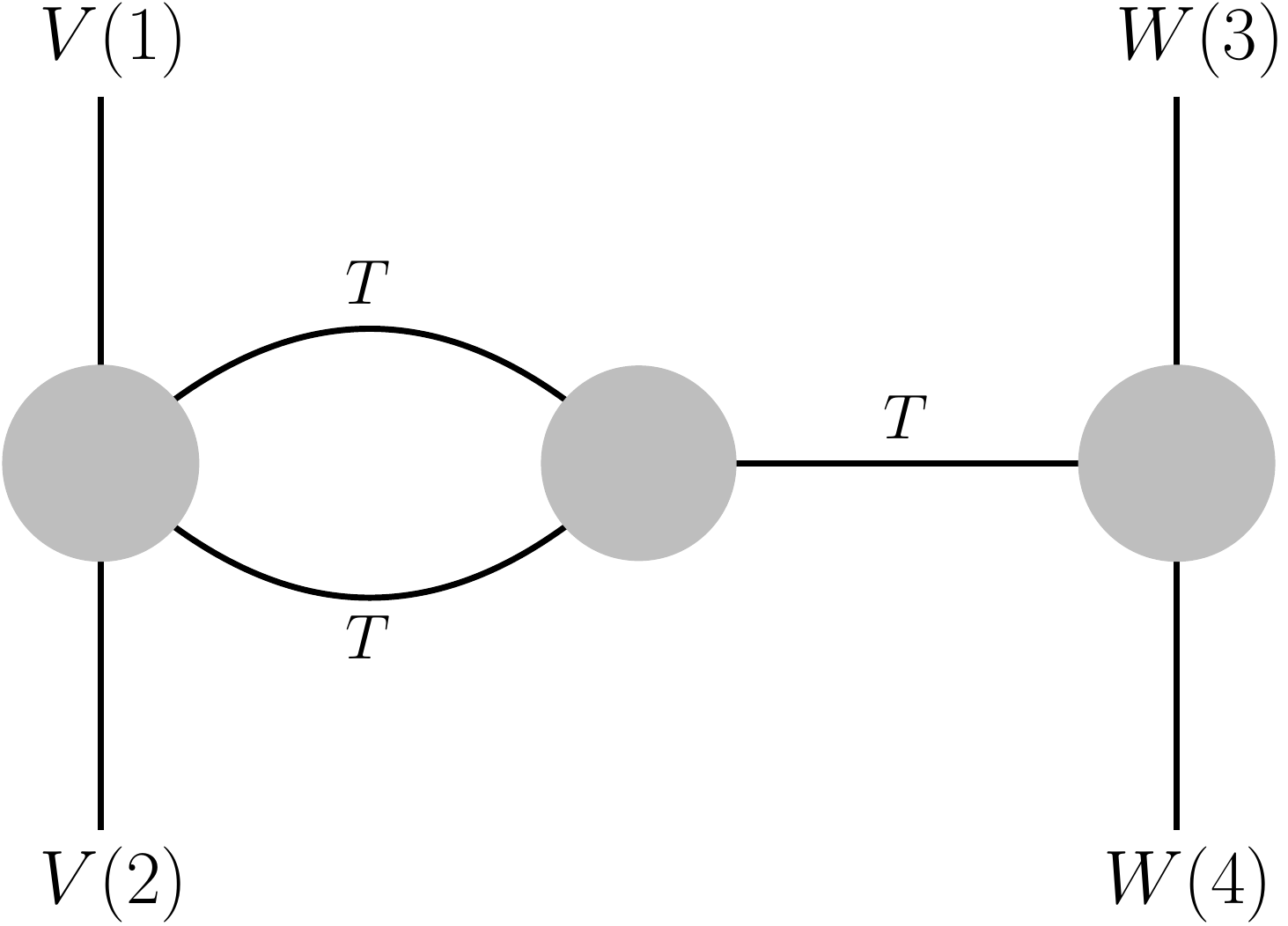}
		\caption{}
		\label{fig:B2B1}
	\end{subfigure}\\
	\vskip 4mm
	\begin{subfigure}[b]{.49\textwidth}
		\centering
		\includegraphics[scale=0.5]{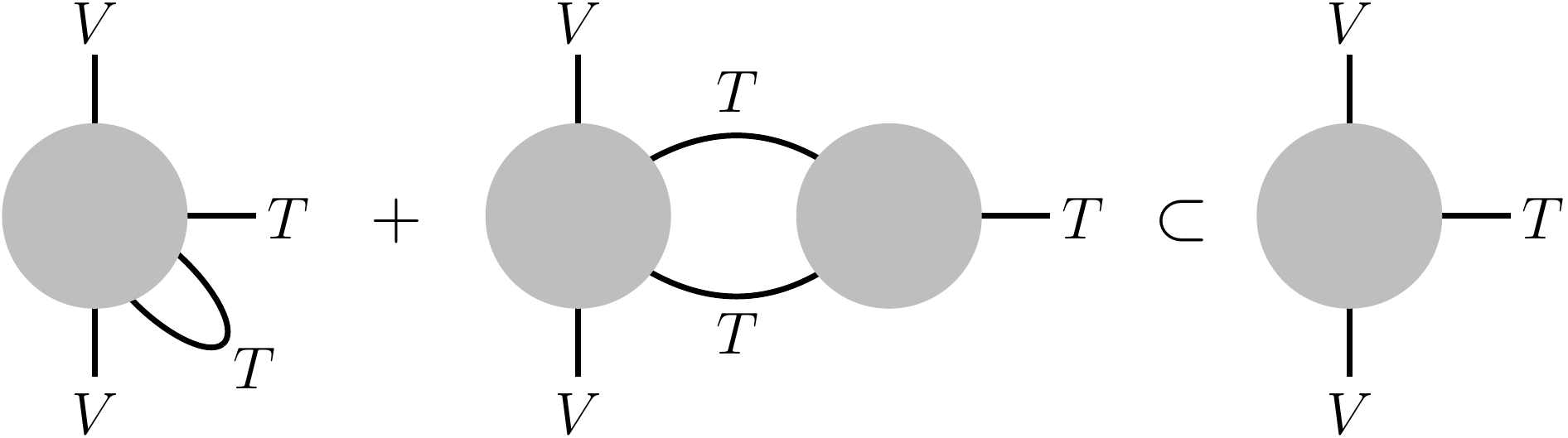}
		\vskip 4mm
		\caption{}
		\label{fig:vertices}
	\end{subfigure}\\
	\caption{(a)-(b) Additional Feynman diagrams that potentially contribute to Virasoro identity blocks at order $O(1/c^2)$. (c) The effect of these diagrams is to renormalize the left vertex of the diagram displayed in Figure~\ref{fig:Feynman1} and occurring at order $O(1/c)$. Since the latter is already exact, one might want to simply discard them.}
	\label{fig:singular diagrams}
\end{figure}

\paragraph{UV divergences and their regularization.} 
As mentioned in section~\ref{section:exchanges}, some terms arising from the central formula \eqref{eq:rule 3} for computing Virasoro identity blocks within the reparametrization formalism suffer from ultraviolet divergences. A general regularization procedure of some sort is needed, which has not been provided so far\footnote{In the heavy-light limit, background subtraction has been successfully applied \cite{Cotler:2018zff}. Away from this limit, there is however no obvious reference background to subtract from.}. A similar issue potentially arises when evaluating Feynman diagrams. At order $O(1/c^2)$ for instance, one can consider the diagrams of Figure~\ref{fig:B3B1}-\ref{fig:B2B1} in addition to that of Figure~\ref{fig:Feynman2}. The appearance of the stress tensor running in loops implies that these also suffer from ultraviolet divergences. One should in fact identify the diagrams displayed in Figure~\ref{fig:B3B1} and Figure~\ref{fig:B2B1} with the terms $\langle \mathcal{B}^{(3)}_{h_V}(1,2) \mathcal{B}^{(1)}_{h_W}(3,4) \rangle_c$ and $\langle \mathcal{B}^{(2)}_{h_V}(1,2) \mathcal{B}^{(1)}_{h_W}(3,4) \rangle_c$ in \eqref{eq:F4 subleading}, respectively. Since both approaches require regularization, a natural strategy would consist in applying one of the textbook regularization procedures to the evaluation of Feynman diagrams and deduce the corresponding rules within the reparametrization formalism. But one could also argue that these diagrams should be discarded altogether on the basis that they seem to correct the vertices appearing in the diagram at order $O(1/c)$ displayed in Figure~\ref{fig:Feynman1}. Since these vertices are already exact as previously emphasized, they may not need to be renormalized. We illustrate this in Figure~\ref{fig:vertices}. This reasoning seems in agreement with the regularization procedure of the gravitational Wilson line formalism \cite{Fitzpatrick:2016mtp}, where one only keeps terms corresponding to stress tensors propagating between both pairs of primary operators. We hope to come back this issue in the future. 

\paragraph{Heavy-light limit.} The reparametrization formalism has also been used to efficiently compute Virasoro identity blocks in the heavy-light limit $h_V=O(1)\,, h_W=O(c)$ \cite{Cotler:2018zff}. It would be interesting to revisit this computation in terms of Feynman diagrams along the lines suggested by the present work. This would in principle require the resummation of a very large number of diagrams. Note that a similar resummation was explicitly performed in \cite{Fitzpatrick:2015foa} using a different approach. However, it is known that the heavy operator insertions can be simulated by an appropriate thermal background \cite{Fitzpatrick:2015zha,Anous:2019yku,Vos:2020clx}. At leading order in the heavy-light limit, a Virasoro identity block reduces to the two-point function of a light operator in a thermal background.  Cotler and Jensen showed that an analogous statement holds within the reparametrization formalism, namely that the heavy-light Virasoro identity block is obtained from the bilocal expectation value $\langle \mathcal{B}_{h_V}(1,2) \rangle_{\text{thermal}}$. In fact, the latter quantity naturally contains subleading corrections associated with $\epsilon$'s running in loops~\cite{Cotler:2018zff}. One could easily set up this computation in terms of Feynman diagrams involving stress tensor exchanges between a single pair of light operators in the appropriate thermal background.    

\paragraph{Gravitational theories and holography.} We end this discussion by mentioning the relevance of reparametrization modes to gravitational theories and holography, which has been our initial motivation to perform the present study. Especially clear is their role within the AdS$_3$/CFT$_2$ correspondence, where it was shown that the gravitational on-shell action coincides with the Liouville version \eqref{eq:Liouville action} of the Polyakov generating functional \cite{Skenderis:1999nb,Manvelyan:2001pv}. This makes perfect sense since the AdS/CFT dictionary precisely identifies the bulk on-shell action with the generating functional of a dual CFT \cite{Gubser:1998bc,Witten:1998qj}. Hamiltonian reductions of three-dimensional gravity with AdS asymptotics were also shown to yield either Liouville theory \cite{Coussaert:1995zp} or the Alekseev--Shatashvili action \cite{Cotler:2018zff}\footnote{The authors of \cite{Cotler:2018zff} further argued that the Alekseev--Shatashivili action may be successfully quantized, leading to a quantum theory of boundary gravitons.}. Here, we gave a unified view of the different forms taken by the generating functional of stress tensor correlations from the perspective of 2d CFTs. The Alekseev--Shatashvili action also appeared from Hamiltonian reductions of three-dimensional gravity with de Sitter asymptotics \cite{Cotler:2019nbi} and of the superrotation sector of four-dimensional gravity with flat asymptotics \cite{Nguyen:2020hot}. We believe that these constitute important hints to the holographic nature of these gravitational theories away from the well-understood and heavily studied AdS/CFT correspondence.  

\section*{Acknowledgments}
I thank Jordan Cotler and Jakob Salzer for useful comments on a draft of this paper, and Jakob Salzer for collaboration on related topics. I also thank Felix Haehl, Gideon Vos and Peter West for interesting discussions. This work is supported by a grant from the Science and Technology Facilities Council (STFC).

\appendix

\section{Conformal Ward identities}
\label{appendix:4 point}
We recall the conformal Ward identity for $n$ stress tensor insertions \cite{Belavin:1984vu}
\begin{gather}
\label{eq:conf Ward}
\langle T(w_1)...T(w_n) \Oh(1) \Oh(2) \rangle\\
\nonumber =\left[\sum_{i=2}^n \left(\frac{2}{w_{1i}^2}+\frac{\partial_{w_i}}{w_{1i}}\right)+\sum_{j=1,2} \left(\frac{h}{(w_1-z_j)^2}+\frac{\partial_{z_j}}{w_1-z_j}\right)\right] \langle T(w_2)...T(w_n) \Oh(1) \Oh(2) \rangle\\
\nonumber
+\sum_{i=2}^n \frac{c/2}{w_{1i}^4}\, \langle T(w_2)...T(w_{i-1})T(w_{i+1})...T(w_n) \Oh(1) \Oh(2) \rangle\,.
\end{gather}
In the case of two stress tensor insertions, it yields
\begin{gather}
\label{eq:TTVV appendix}
\langle \partialbar T(w_1) \partialbar T(w_2)\Oh(1)\Oh(2) \rangle\\
\nonumber
=(2\pi)^2 \langle \Oh(1) \Oh(2) \rangle \left[C^h_{w_1}(1,2)C^h_{w_2}(1,2)+C^h_{\lbrace w_1,w_2\rbrace}(1,2)\right]\,,
\end{gather}
with 
\begin{gather}
C_w^h(1,2)= -h \left[\left(\partial_w +\frac{2}{z_{12}}\right) \delta(w-z_1)+(z_1 \leftrightarrow z_2)\right]\,,\\
\label{eq:Cww}
C^h_{w_1,w_2}(1,2)=2C^h_{w_1}(1,2) \partial_{w_1}\delta(w_{12}) +\delta(w_{12}) \partial_{w_1} C^h_{w_1}(1,2)\\
\nonumber +h\big[\delta(w_2-z_1)\left(\partial_{w_1}^2+\frac{2\partial_{w_1}}{z_{12}}+\frac{2}{z_{12}^2}\right)\delta(w_1-z_1)-\frac{2\delta(w_2-z_1)\delta(w_1-z_2)}{z_{12}^2}+(z_1 \leftrightarrow z_2)\big]\,.
\end{gather}
The symmetrizer $ \lbrace w_1,w_2\rbrace=\frac{1}{2} \left[(w_1, w_2)+(w_2, w_1)\right]$ ensures that \eqref{eq:TTVV appendix} is symmetrical under $w_1 \leftrightarrow w_2$ as it should. When symmetrized, the terms on the first line of \eqref{eq:Cww} actually cancel as distributions. Indeed, integrating against a test function $f(w_1,w_2)$ with compact support, we have
\begin{subequations}
\begin{align}
&\int d^2w_1 d^2w_2\, f(w_1,w_2)\left[ C^h_{w_1} \partial_{w_1} \delta(w_{12})+C^h_{w_2} \partial_{w_2} \delta(w_{12})+\delta(w_{12}) \partial_{w_1} C^h_{w_1}(1,2)\right]\\
&=-\int d^2w \left[f^{(1,0)}(w,w) C^h_w+f^{(0,1)}(w,w) C^h_w+f(w,w)\partial_w C^h_w \right]\\
&=-\int d^2w\, \frac{d}{dw}\left(f(w,w)C^h_w \right)=0\,.
\end{align}
\end{subequations}
The careful reader already sees the structure of the reparametrization formalism appearing at this stage. Indeed, integration of $C^h_w(1,2)$ and $C^h_{\lbrace w_1,w_2 \rbrace}(1,2)$ against $\epsilon$ propagators yields the bilocal vertex operators $b_h^{(1)}(1,2)$ and $b^{(2)}_h(1,2)$, respectively.

At large central charge and in the light-light limit $h=O(\sqrt{c})$, repetitive use of the conformal Ward identity \eqref{eq:conf Ward} yields 
\begin{gather}
\label{eq:TTTVV appendix}
\langle \partialbar T(w_1)...\partialbar T(w_n) \Oh(1)\Oh(2) \rangle\\
\nonumber
=(2\pi)^n \langle \Oh(1)\Oh(2) \rangle \left[C^h_{w_1}(1,2)...C^h_{w_n}(1,2)+O(1/\sqrt{c})\right]\,.
\end{gather}

\bibliography{bibl}
\bibliographystyle{JHEP}
\end{document}